\documentclass[useAMS,usenatbib]{mn2e}
\usepackage{epsfig,amsmath,natbib}
\usepackage{color}

\def\be{\begin{equation}} 
\def\ee{\end{equation}} 
\def\ba{\begin{eqnarray}} 
\def\ea{\end{eqnarray}}

\def\cc{\,{\rm {cm^{-3}}}}

\def\gsim{\lower.5ex\hbox{\gtsima}} 
\def\lsim{\lower.5ex\hbox{\ltsima}} \def\gtsima{$\; \buildrel > \over 
\sim \;$} \def\ltsima{$\; \buildrel < \over \sim \;$} \def\prosima{$\; 
\buildrel \propto \over \sim \;$} \def\gsim{\lower.5ex\hbox{\gtsima}} 
\def\lsim{\lower.5ex\hbox{\ltsima}} 
\def\simgt{\lower.5ex\hbox{\gtsima}} 
\def\simlt{\lower.5ex\hbox{\ltsima}} 
\def\simpr{\lower.5ex\hbox{\prosima}}   
  
 \def\gtsima{$\; \buildrel > \over \sim \;$} 
\def\ltsima{$\; \buildrel < \over \sim \;$} 
\def\gsim{\lower.5ex\hbox{\gtsima}} 
\def\lsim{\lower.5ex\hbox{\ltsima}} 
\def\simgt{\lower.5ex\hbox{\gtsima}} 
\def\simlt{\lower.5ex\hbox{\ltsima}} 
\def\simpr{\lower.5ex\hbox{\prosima}}

\def\E3{{\cal E}_{\rm g}^{III}}

\def\r12{r_{1/2}} 
\def\x12{x_{1/2}} 
\def\v12{v_{1/2}}

\def\Tcmb{T_{\textrm{CMB}}}

\title[Dust growth in galaxies]{The problematic growth of dust in high redshift galaxies} 
\author[Ferrara et al.]{A. Ferrara$^{1}$, S. Viti$^{2}$, C. Ceccarelli$^{3,4}$  \\
$^{1}$ Scuola Normale Superiore, Piazza dei Cavalieri 7, I-56126 Pisa, Italy\\
$^{2}$ Department of Physics and Astronomy, University College London,
Gower Street, London, WC1E 6BT, United Kingdom \\
$^{3}$ Univ. Grenoble Alpes, IPAG, F-38000 Grenoble, France \\
$^{4}$ CNRS, IPAG, F-38000 Grenoble, France \\}
\begin{document} 
 
\date{\today} 
 
\pagerange{\pageref{firstpage}--\pageref{lastpage}} \pubyear{2016} 
 
\maketitle 
 
\label{firstpage} 
\begin{abstract} 
Dust growth via accretion of gas species has been proposed as the dominant process to increase the amount of dust in galaxies. We show here that this hypothesis encounters severe difficulties that make it unfit to explain the observed UV and IR properties of such systems, particularly at high redshifts. Dust growth in the diffuse ISM phases is hampered by (a) too slow accretion rates; (b) too high dust temperatures, and (c) the Coulomb barrier that effectively blocks accretion. In molecular clouds these problems are largely alleviated. Grains are cold (but not colder than the CMB temperature, $\Tcmb \approx 20$ K at redshift $z=6$). However, in dense environments accreted materials form icy water mantles, perhaps with impurities. Mantles are immediately  ($\simlt 1$ yr) photo-desorbed  as grains return to the diffuse ISM at the end of the cloud lifetime, thus erasing any memory of the growth. We conclude that dust attenuating stellar light at high-$z$ must be ready-made stardust largely produced in supernova ejecta. 
\end{abstract}


\section{Introduction}
\label{Mot}

Dust grains are a fundamental constituent of the interstellar medium (ISM) of galaxies. A large fraction ($\approx $ 50\% in the Milky Way) of the heavy elements produced by nucleosynthetic processes in stellar interiors can be locked into these solid particles. They are vital elements of the ISM multiphase gas life-cycle, and key species for star formation, as they absorb interstellar UV photons, heat and cool the gas- They also catalyze the formation of H$_2$ on their surfaces, the first step toward the formation of all other ISM molecules, including CO.

The presence of dust at high ($z\simgt 6$) redshift implies that
conventional dust sources (AGB and evolved stars) are not the dominant
contributors. This is because their evolutionary timescales are close
or exceed the Hubble time at that epoch ($\approx 1$ Gyr). Following
the original proposal by \cite{Todini01}, it is now believed that the
first cosmic dust grains were formed in the supernova ejecta ending
the evolution of fast-evolving massive stars \citep{Hirashita02,Nozawa07,
  Bianchi07, Gall11, Bocchio16}. Thus, albeit quasar host galaxies show remarkably high dust masses \citep{Bertoldi03, Beelen06, Michalowski10}, in general the dust content of early galaxies rapidly decreases \citep{Capak15, Bouwens16}. This does not come as a complete surprise given that the mean metallicity of the Universe\footnote{Throughout the paper,  we assume a flat Universe with the following cosmological parameters:  $\Omega_{\rm m} = 0.308$, $\Omega_{\Lambda} = 1- \Omega_{\rm m} = 0.692$, and $\Omega_{\rm b} = 0.048$,  where $\Omega_{\rm M}$, $\Omega_{\Lambda}$, $\Omega_{\rm b}$ are the total matter, vacuum, and baryonic densities, in units of the critical density,   and  $h$ is the Hubble constant in units of 100 km/s \citep{Ade15}.}  increases with time.

Usually the presence of dust in high-$z$ galaxies is assessed via a specific observable, the so-called $\beta$ slope. This is defined as
the slope of the rest-frame UV galaxy emission spectrum, $F_\lambda^i \propto \lambda^{\beta}$, in the wavelength range $1600-2500$ A. As dust extinction typically rises towards shorter wavelengths, a flatter slope indicates the presence of larger amounts of dust. Indeed this is what has been recently shown by ALMA observations \citep{Capak15,Bouwens16}. 

It has been claimed that current observations cannot be explained purely by dust production by sources (either SNe and AGB/evolved stars). Instead, the dominant contribution to the dust mass of high-$z$ galaxies should come from grain growth \citep{Michalowski10, Hirashita14a, Mancini15} in the interstellar medium. This can happen only if gas-phase atoms and molecules can stick permanently to grain surfaces (typically consisting of silicate or amorphous carbonaceous materials)  and remain bound to the grain solid structure. 
Simplistic approaches based on a ``sticking coefficient'' argument predict that the growth time of the grains could be very fast
($\simlt 1$ Myr). However, such conclusion fails to catch some critical points that we examine here.

The first problem is linked to the type of material that silicate or carbonaceous cores tend to accrete. As we will argue  in this
\textit{Letter}, accretion can only occur in MCs. There, silicate or carbonaceous refractory cores must predominantly accrete the
most abundant elements, namely oxygen and carbon, either in atomic or molecular form (CO). These species condense on grain surfaces where they are hydrogenated, forming icy mantles mainly made up of water, CO, CO$_2$, ammonia, methane and, possibly, other impurities \citep{Boogert15}. Mantles do not share the same optical properties as refractory cores. However, this is not a relevant point, given that icy mantles are very volatile and rapidly photo-desorbed as grains re-emerge from MCs. 

A second problem with the growth scenario arises from the increasing CMB temperature, $\Tcmb= T_0 (1+z)$ K with $T_0=2.725$. While dust in Milky Way molecular clouds typically attains temperatures $\approx 10-20$ K \citep{Stutz10}, at high-$z$ dust cannot cool below $\Tcmb$. This might represent a serious problem for dust growth, as the warmer surface hampers the sticking ability of particles. 

It then appears that if early galaxy properties require a larger amount of UV-absorbing dust particles, we are forced to conclude that the observed grains must be ready-made products of the most efficient high-$z$ factories, i.e. SNe. In the following we show that this is indeed the case.
\section{Dust accretion: what and where}
\label{Met}
After solid particles condense out of early supernova ejecta, and are possibly processed by the reverse shock, they are injected in the
pervasive diffuse phases of the interstellar medium, i.e. the Cold Neutral Medium (CNM, density $n_C \approx 30\cc$, temperature
$T_C\approx 100 $K) and the Warm Neutral Medium (WNM, $n_W \approx 0.4\cc$, $T_W\approx 8000$K). Usually these two phases are considered to be in thermal equilibrium \citep{Field69, Wolfire03, Vallini13}. However, the multiphase regime exists only in a narrow, metallicity-dependent range of pressures. Outside this regime, the gas settles on a single phase. At high pressures (a more typical situation for denser high-$z$ galaxies), the CNM dominates; at low pressures the WNM takes over. We therefore concentrate in what follows on the CNM.  As both destruction and production processes ultimately rely on supernova explosions \citep{Dayal10}, the equilibrium dust abundance is essentially independent on star formation rate (SFR). 

Even if they initially reside in the CNM, grains become periodically embedded into MCs forming out of this phase. This happens on the gas depletion timescale, $\tau_g \approx M_g/\mathrm{SFR}$, where $M_g$ is the total gas mass. In galaxies, $\tau_g$ decreases with redshift, but it remains close to 10\% of the Hubble time at any epoch, i.e. $\tau_g \approx \mathrm{few} \times 100$ Myr at $z=6$.  This timescale should be compared with the lifetime of MCs, which is much shorter, i.e. $\approx 10$ Myr. Thus, it is reasonable to conclude that grains spend a large fraction of their lifetime in the diffuse phase. An important point is that, while in MCs, dust grains are largely shielded from UV radiation and therefore have lower temperatures.

\subsection{The grain accretion process}
Before analysing the specific cases of the CNM and MCs in high-$z$ galaxies, we recall here the basic theory of grain accretion.
Consider grains whose temperature is $T_d$. The grain accretion time scale, $\tau_a$, is set by kinetics and is given by \citep{Spitzer78, Umebayashi80}:
\ba
\tau_a^{-1} \approx S n_d \pi a^2 v_s       
\label{eq1a}
\ea
where $n_d$ is the grain number density (proportional to the hydrogen
number density, $n_H$, and metallicity, $Z$); $a$ is the average
(spherical) grain radius. The most probable species velocity (the
maximum of the Maxwell distribution function) is
$v_s = (2 k_B T/m_s)^{1/2}$, i.e. the square root of the ratio of gas
temperature and species mass; $S(T_d, m_s)$ is the accreting
species sticking probability.  The latter is a poorly known function
of the dust temperature and composition, and of the accreting
species. Recently, \cite{He16} reported an experimental determination
of the $S$ value for different species and dust temperatures. They
propose a general formula to evaluate $S$ from the species binding
energy and dust temperature. In the following, we adopt their
prescription (namely their Eq. 1). In practice, though, for
$T_d \simlt 30$ K, and binding energies larger than about
$E_b/k_B =1100$ K, the sticking probability is equal to unity.

Once an atom or molecule has sticked onto the dust grain, its fate depends on the species binding energy, $E_b$, the dust temperature, $T_d$, and the irradiation from FUV and cosmic rays. First, the atom/molecule might be thermally desorbed after a time 
\ba
  \tau_d^{-1} \approx \nu_0 \exp(-E_b/k_B T_d)
  \label{eq1b}
\ea 
where $\nu_0$ is the vibrational frequency of the sticking species. In general, $\nu_0$ 
depends on the properties of the grain surface and adsorbed species
\citep{Hasegawa92}. In practice, it is $\approx 10^{12} \rm s^{-1}$
for the cases relevant to the present study, namely H, O and Si atoms
and water molecules \citep{Minissale16}.
The binding energy refers to the van der Waals force, typically a small fraction of eV for many species (see below).
Therefore, the desorption rate is very sensitive to $T_d$ and it is almost a step function
\citep{Collings04}. In addition to thermal desorption, FUV and cosmic
rays may provide enough energy to the accreted species to evict a
fraction of them back into the gas \citep{Leger85, Shen04,
  Bertin13}. This desorption rate depends on $E_b$ and the specific
microphysics of the process. It suffices to note here that
Eq. \ref{eq1a} provides, therefore, a lower limit to the actual
accretion time.

Finally, a third relevant timescale is that needed by a species to hop and scan the grain surface. This is known as the ``scanning time'', and it is given by:
\ba
  \tau_s^{-1} \approx N_s^{-1} \nu_0 \exp(-E_d/k_BT_d).
  \label{eq1c}
\ea
$N_s$ is the number of sites of the grain; for $a=0.1 \mu$m and a mean
distance between sites of 3.5 A, $N_s\approx 10^5$.
The diffusion energy, $E_d$, is a poorly known parameter. Usually,
this is taken to be a fraction, $f_d$, of the binding energy,
i.e. $E_d = f_dE_b$. Experiments give often rather contradictory
results, but tend to justify values in the range $f_d=0.3-0.8$ (see
discussion in \cite{Taquet12}).  For highly reactive species, like H and O, laboratory experiments provide diffusion energies rather than $E_b$. Fox oxygen, \cite{Minissale16} find $E_d/k_B =750$ K and $E_b/k_B =1320$ K. For H atoms, experiments find $210\, \mathrm{K} < E_d/k_B < 638$ K, depending on adsorbing surface and site \citep{Hornekaer05, Matar08, Hama12}. However, \citet{Hama12} showed that the majority of sites have $E_d/k_B= 210$ K. In this study, we assume $E_b/k_B=500$ K and $E_d/k_B= 210$ K for H, following the majority of astrochemical models. No experimental measurements nor theoretical calculations exist for the binding energy of Si atoms, apart from the 2700 K heuristic estimate by \cite{Hasegawa93}. Finally, the water binding energy has been measured to be 1870 and 5775 K on bare silicates/amorphous water, respectively \citep{Fraser01}.

In the following we investigate whether grain growth is possible and what type of condensables can be actually accreted on the bare soil surfaces of the grains. We consider separately the case of the CNM and MC environments and study processes as a function of the galaxy redshift. We will also concentrate on silicates as \cite{Weingartner01} showed that a for SMC-like extinction curve appropriate for high-$z$ galaxies, the silicate/carbonaceous mass ratio is $\approx 11:1$, i.e. C-based grains negligibly contribute to total dust mass.

\subsection{Residence in the diffuse phase} 

Let us consider the case of the CNM, whose properties have been defined above, namely a gas temperature of 100 K and hydrogen nuclei density of 30 cm$^{-3}$. Assume also that the sticking species is a Si atom (of mass 28 amu). Plugging these values into Eq. \ref{eq1a} we find 

\be 
\tau_a \approx 1.2 \, \textrm{Gyr}  %
\left({0.1 Z_\odot\over Z }\right) %
\left({a \over {0.1 \mu \textrm{m}}}\right) %
S^{-1}
\label{eq1}
\ee
The binding energy of Si atoms is estimated to be $E_b/k_B=2700$ K
(Hasegawa \& Herbst 1993; see above), so that the sticking probability
is essentially unity, according to \cite{He16}.  This time scale is
very long and even exceeds the Hubble time at $z=6$. Smaller grains,
$a\approx 0.01 \mu$m, due to their larger surface/mass area, might
have proportionally shorter $\tau_a$. However, times scales remain
comparable or longer than the depletion time scale, $\tau_g$,
discussed above.

Thus, grain growth in the diffuse ISM is overwhelmingly difficult for at least three reasons. First, the accretion time scale $\tau_a$ is at best comparable to the residence time in the ISM. This means that by the time at which accretion from the gas phase \textit{might} become important, the grain moves from the diffuse to the dense molecular phase. However, the situation is worse than this due to two additional complications.

Grain temperatures in the CNM are rather high (at least compared to that in MCs). Due to their compactness (sizes $\simlt 1$ kpc) and
consequently high surface star formation rate per unit area ($\dot \Sigma_*\approx 1 M_\odot$ yr$^{-1}$ kpc$^{-2}$ -- about 100
times that of the Milky Way), the interstellar UV field is correspondingly more intense. Roughly scaling the value of the Habing
flux with $\dot \Sigma_*$ gives a value of $G_0 \approx 100$. Grains achieve temperatures ranging from 30 to 50 K for a typical
$a=0.1\, \mu$m grain. Smaller grains, which provide most of the surface area for accretion, are even hotter, $T_d=42-72$ K (\cite{Bouwens16}, Ferrara \& Hirashita, in prep.). Under these conditions, Si,  and O atoms will not remain attached to the grain surface. Instead, they almost instantaneously (fraction of a second) bounce back into the gas  (see Eq. \ref{eq1b}).

Further complications arise from the grain charge. Under the action of strong UV irradiation, grains attain a positive charge. For example, \cite{Bakes94} show that a spherical grain located in the CNM and exposed to a UV flux of intensity $G_0=1$ (in standard Habing units) attains an equilibrium charge $Z_g \approx +10$.  This value must be seen as a lower limit as we mentioned already that $G_0$ can be up to a factor 100 higher in early systems.  This represents a problem for accretion of key condensable species as, e.g. Si and C. Their ionization potential (11.26 eV for C and 8.15 eV for Si) is lower than 1 Ryd. Thus, even in neutral regions as the CNM these species are ionized. As a result, the Coulomb repulsion between the charged grains and these ions represents a virtually unsurmountable barrier in order for these species to reach the grain surface. 

These processes acting against accretion of materials from the surrounding gas by grains lead to conclude that, under the
conditions prevailing in high-$z$ galaxies, grain growth in the diffuse ISM phases can be safely excluded. We now turn to the analysis of the MC case.

\begin{figure*}
\center
\includegraphics[width=90mm]{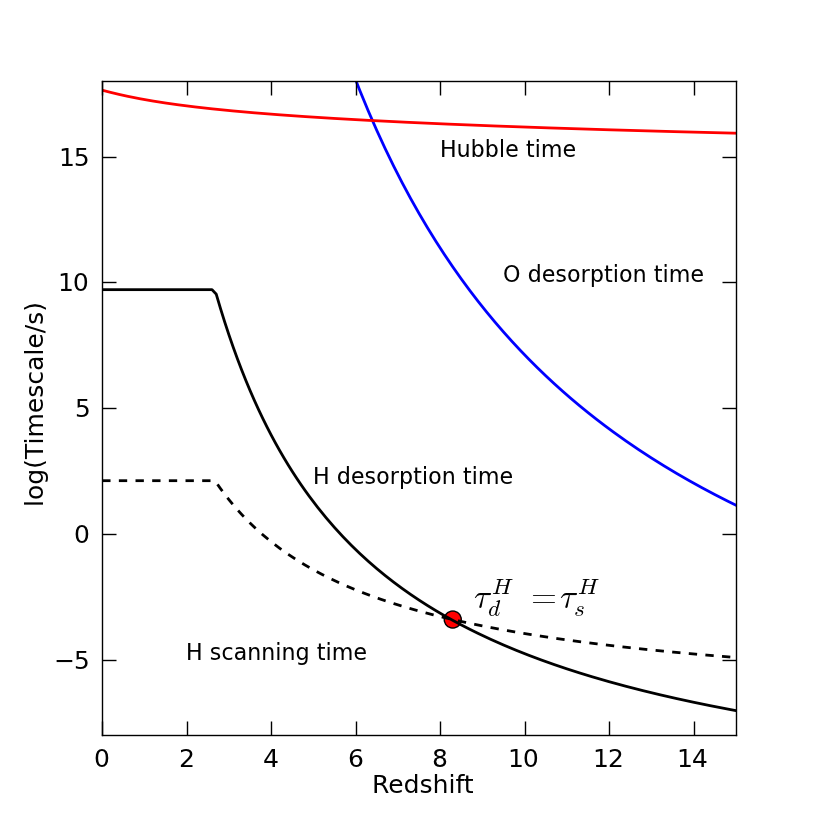}
\caption{Comparison among the different relevant times scales for ice formation on grain surfaces. The H-atom (solid line) desorption time, $\tau_d^H$, is compared against the H scanning time (dashed) of the grain surface, $\tau_s^H$. Ice formation is possible only when $\tau_s^H < \tau_d^H$, i.e. at all redshifts $z< 8.3$. A pre-requisite is that the analogous oxygen desorption time (blue solid curve) is larger than both previous timescales to ensure that oxygen atoms are always available to react and form H$_2$O. Note that the O desorption time is very long and also larger than the Hubble time (red solid curve) for $z \simlt 6.5$. We have assumed a grain temperature $T_d = \max(10\, \mathrm{K}, \Tcmb)$.  
} 
\label{Fig01}
\end{figure*}

\subsection{Residence in molecular clouds}

Accretion conditions become apparently more favorable when grains are incorporated into a newly born MC. There, due to the higher mean gas density ($n=10^{3-4} \cc$) the gas accretion timescale becomes shorter. In spite of the lower temperatures ($\approx 10$ K), the large density found in MCs boosts the accretion rate with respect to the diffuse ISM by a factor $\approx 10-100$.  While residing in MCs, dust grains can evolve mostly because of the growth of icy mantles coating the grain refractory cores (e.g. \cite{Caselli12, Boogert15}.

Due to the attenuation or total suppression of the UV field, grains are colder in MCs. However, at high redshift heating by CMB
becomes important and sets the minimum temperature of gas and grains in MCs. This value depends on redshift, and at $z\geq 6$ it is $T_d \geq 19$ K. This is about a factor of two larger than MC temperatures in the local universe. At these warmer temperatures, processes like thermal desorption (see Eq. \ref{eq1b}), may dominate (or even impede) grain growth.

In the case of MW dust, a grain embedded in MCs is coated with several ($\approx 100$) layers of water ice \citep{Taquet12}.  This is because oxygen, the most abundant element after H and He, after landing on the grain surface, is rapidly\footnote{When the dust temperature is 10 K, the permanence time of H atoms on the grain surface is $\approx 165$ yr, while it takes only $\approx 132$ s for each H atom to scan the full grain surface (Fig. \ref{Fig01}).} hydrogenated, and forms water molecules \citep{Dulieu10}. Water remains then stuck on the grain surface. Besides being the most abundant element, O condenses also faster than other heavier elements, like Si or SiO (which are particularly relevant for this study), according to Eq. \ref{eq1a}.

The water formation process on grains is a two-step process\footnote{Here we neglect a number of subtleties and complications related to water formation, which does not involve only the simple direct addition of H and O atoms \citep{Dulieu10}.}. First, an impinging oxygen atom must become bound to the surface. Second it must react with H atoms landed on the grain on a time scale shorter than their desorption time scale, $\tau_d$, during which they scan the grain surface. As dust and gas temperatures are likely higher in MCs at high redshifts due to CMB, we need to re-consider the efficiency of this process, in terms of the above timescales.

Consider the formation of the first H$_2$O layer. The desorption (or
permanence on the grain surface) time (Eq. \ref{eq1b}) for O atoms is
larger than the Hubble time for $z \simlt 6.5$ (see Fig. \ref{Fig01}).
H atoms permanence time depends on $z$ (Fig. \ref{Fig01}) and it is
$\approx 240$ ms at $z=6$.  These timescales have to be compared with
the time necessary for H atoms to scan the grain surface, $\tau_s$
(Eq. \ref{eq1c}). Taking again $a=0.1\, \mu$m, and $N_s = 10^5$ sites to scan,
it takes $\tau_s =6$ ms for H atoms to scan a grain at temperature
$T_d=\Tcmb(z=6)$, and form H$_2$O molecules. Hence, the time required to form the first water layer is set by the accretion rate of oxygen atoms. Assuming a MC hydrogen density  $n_H=10^4 \cc$, a temperature  $\Tcmb(z=6)$, and that a layer contains $N_s \approx 10^5$ molecules, we get that the first layer is formed in $\approx 1500$ yr.  Once formed, water (whose
$E_b$ is larger than the O atom binding energy) will remain
frozen forming a mantle.  The key point here is that, once a first
layer of water ice is formed, this will prevent Si atoms or SiO molecules to get in contact with the silicate surface. On the
contrary, given the much smaller Si abundance (for solar ratios Si/O $\approx 1/200$ by number), these species will be -- at best, trapped in the ice layer, forming impurities but not silicate-like
bonds\footnote{Note that the seminal and important experimental
  results by \cite{Krasnokutski14} suggesting the formation of
  silicate-like bonds from accreting Si and SiO molecules, refer to
  ultra-low temperatures ($\sim 0.34$ K) where quantum chemical
  effects might largely enhance the efficiency of the reaction. This
  is caused by trapping of the very cold molecules in a van der Waals
  energy potential which is easily overcome at higher temperatures
  (see, for example, a conceptually similar experiment by
  \cite{Shannon13}).}. As a final remark, we underline that the above
rapid hydrogenation process applies also to the case of carbonaceous
grains.

\subsection{Mantle photo-desorption in the diffuse ISM}
\label{Met}
The results of the previous subsection lead us to conclude that grain growth in molecular clouds is largely in the form of icy mantles. 
Once the parent molecular cloud is dispersed (typical lifetime 10 Myr) as a result of radiative and mechanical feedback from stars born in its interior, core-mantle grains are returned to the ISM. However, the memory of growth occurred in the MC will be promptly erased as icy mantles are photo-desorbed by the FUV field  \citep{Barlow78, Fayolle13}. A simple calculation shows that this timescale is very short. Suppose that the ice mantle is made by $\ell = 100$ layers, each containing $N_s=10^5$ sites. The time necessary to completely photo-desorb the mantle, $\tau_p$, is that required to provide each site with an UV photon. Thus, for a given Habing band ($6-13.6$ eV) UV flux intensity, $G_0$ (in cgs units), and mean photon energy $\langle h\nu \rangle \approx 10$ eV 
\be
y_{H_2O}{G_0\over \langle h\nu \rangle} \pi a^2 \tau_p = N_s \ell,
\ee  
or $t_p = 10-1000$ yr for $G_0=100-1$, respectively. In the previous
calculation we have further assumed a H$_2$O photo-desorption yield, $y_{H_2O}=0.001$, following \cite{Oberg09}.   To all purposes the accreted mantle material (along with the impurities) is almost instantaneously  ``lifted'' from the core. The bare grain then finds itself back in the ISM with (optical) properties essentially identical to those prevailing before the journey into the MC, and set by nucleation processes in sources.

\section{Conclusions}
\label{Con}
We have shown that grain growth by accretion is a problematic process hampered by a number of difficulties which become additionally more severe at high redshifts as the temperature floor set by the CMB increases (i.e. grains are hotter). In the CNM (for the WNM the situation is even less favorable) grain growth is essentially prevented by the (i) low accretion rates related to low densities; (b) higher dust temperatures (particularly in compact, high-$z$ galaxies) causing very short thermal desorption times; (c) Coulomb repulsive forces preventing positively charged ions (Si, C) to reach the grain surface. 

Molecular clouds offer more favorable conditions, due to their high density and low dust temperature. However, we have shown that once the bare grain cores are immersed in the MC environment, they are covered by a water ice mantle on a timescale of a few thousand years at $z=6$.  As hydrogenation is very fast ($\approx$ ms), such timescale is set by the accretion rate of oxygen atoms.  This process fully operates in spite of CMB  heating up to $z \approx 8.3$. Beyond that epoch, predictions become uncertain due to the lack of a precise knowledge of the diffusion energies of the various species. As the grains are returned to the diffuse phase at the end of the MC lifetime, the mantles are photo-desorbed and the bare core is exposed again. That is, the memory of the ice growth phase in the MC is completely erased. 

If grain growth is as problematic as we point out, it is necessary to re-evaluate the arguments, often made, invoking it. These are generically based on a comparison between the dust production rate by sources (planetary nebulae, evolved stars, and SNe) and destruction rate in supernova shocks \citep{Draine09}. According to such, admittedly uncertain, estimates \textit{made for the Milky Way}, dust production fails by about a factor 10 to account for the observed dust mass once shock destruction is accounted for. 

In the light of the present results, reconciliation of this
discrepancy must necessarily come from either (a) an upward revision
of the production rate, or (b) downward reappraisal of the dust
destruction efficiency by shocks. Interestingly, there appears to be
evidence for both solutions, and perhaps even a combination of the
two. Recent studies \citep{Matsuura11, Indebetouw14} have determined
the dust mass produced by SN1987A. The observations revealed the
presence of a population of cold dust grains with $T_d=17-23$ K; the
emission implies a dust mass of $\approx 0.4-0.7\, M_\odot$. Such
value is $20-35$ times larger than what usually assumed in the above
argument. Moreover, \cite{Gall14} found that the $0.1-0.5$ masses of
dust detected in the luminous SN2010jl are made of $\mu$m-size grains
which easily resist destruction in (reverse) shocks. Alternatively, the puzzle can be solved also by decreasing the destruction rate. Indeed, \cite{Jones11} thoroughly reanalyzed this issue and found that the destruction efficiencies might have been severely overestimated. They additionally conclude that ``the current estimates of “global” dust lifetimes could be uncertain by factors large enough to call into question their usefulness''.  Given the situation, the results presented here do not seem to create any additional challenge. On the contrary, the hope is that they will stimulate deeper studies on the key problem of dust evolution in a cosmological context. 

\section*{Acknowledgments} 
This research was supported in part by the National Science Foundation under Grant No. NSF PHY11-25915.

\bibliographystyle{apj}
\bibliography{ref}

\begin{thebibliography}{49}
\expandafter\ifx\csname natexlab\endcsname\relax\def\natexlab#1{#1}\fi

\bibitem[{{Bakes} \& {Tielens}(1994)}]{Bakes94}
{Bakes}, E.~L.~O., \& {Tielens}, A.~G.~G.~M. 1994, \apj, 427, 822

\bibitem[{{Barlow}(1978)}]{Barlow78}
{Barlow}, M.~J. 1978, \mnras, 183, 417

\bibitem[{{Beelen} {et~al.}(2006){Beelen}, {Cox}, {Benford}, {Dowell},
  {Kov{\'a}cs}, {Bertoldi}, {Omont}, \& {Carilli}}]{Beelen06}
{Beelen}, A., {Cox}, P., {Benford}, D.~J., {Dowell}, C.~D., {Kov{\'a}cs}, A.,
  {Bertoldi}, F., {Omont}, A., \& {Carilli}, C.~L. 2006, \apj, 642, 694

\bibitem[{{Bertin} {et~al.}(2013){Bertin}, {Fayolle}, {Romanzin}, \& {Poderoso}
  H.~A.~M.}]{Bertin13}
{Bertin}, M., {Fayolle}, E.~C., {Romanzin}, C., \& {Poderoso} H.~A.~M., e.~a.
  2013, \apj, 779, 120

\bibitem[{{Bertoldi} {et~al.}(2003){Bertoldi}, {Carilli}, {Cox}, {Fan},
  {Strauss}, {Beelen}, {Omont}, \& {Zylka}}]{Bertoldi03}
{Bertoldi}, F., {Carilli}, C.~L., {Cox}, P., {Fan}, X., {Strauss}, M.~A.,
  {Beelen}, A., {Omont}, A., \& {Zylka}, R. 2003, \aap, 406, L55

\bibitem[{{Bianchi} \& {Schneider}(2007)}]{Bianchi07}
{Bianchi}, S., \& {Schneider}, R. 2007, \mnras, 378, 973

\bibitem[{{Bocchio} {et~al.}(2016){Bocchio}, {Marassi}, {Schneider}, {Bianchi},
  {Limongi}, \& {Chieffi}}]{Bocchio16}
{Bocchio}, M., {Marassi}, S., {Schneider}, R., {Bianchi}, S., {Limongi}, M., \&
  {Chieffi}, A. 2016, \aap, 587, A157

\bibitem[{{Boogert} {et~al.}(2015){Boogert}, {Gerakines}, \&
  {Whittet}}]{Boogert15}
{Boogert}, A.~C.~A., {Gerakines}, P.~A., \& {Whittet}, D.~C.~B. 2015, \araa,
  53, 541

\bibitem[{{Bouwens} {et~al.}(2016){Bouwens}, {Aravena}, {Decarli}, {Walter},
  {da Cunha}, {Labbe}, {Bauer}, {Bertoldi}, {Carilli}, {Chapman}, {Daddi},
  {Hodge}, {Ivison}, {Karim}, {Le Fevre}, {Magnelli}, {Ota}, {Riechers},
  {Smail}, {van der Werf}, {Weiss}, {Cox}, {Elbaz}, {Gonzalez-Lopez},
  {Infante}, {Oesch}, {Wagg}, \& {Wilkins}}]{Bouwens16}
{Bouwens}, R., {et~al.} 2016, ArXiv e-prints

\bibitem[{{Capak} {et~al.}(2015){Capak}, {Carilli}, {Jones}, {Casey},
  {Riechers}, {Sheth}, {Carollo}, {Ilbert}, {Karim}, {Lefevre}, {Lilly},
  {Scoville}, {Smolcic}, \& {Yan}}]{Capak15}
{Capak}, P.~L., {et~al.} 2015, \nat, 522, 455

\bibitem[{{Caselli} \& {Ceccarelli}(2012)}]{Caselli12}
{Caselli}, P., \& {Ceccarelli}, C. 2012, \aapr, 20, 56

\bibitem[{{Collings} {et~al.}(2004){Collings}, {Anderson}, {Chen}, {Dever},
  {Viti}, {Williams}, \& {McCoustra}}]{Collings04}
{Collings}, M.~P., {Anderson}, M.~A., {Chen}, R., {Dever}, J.~W., {Viti}, S.,
  {Williams}, D.~A., \& {McCoustra}, M.~R.~S. 2004, \mnras, 354, 1133

\bibitem[{{Dayal} {et~al.}(2010){Dayal}, {Hirashita}, \& {Ferrara}}]{Dayal10}
{Dayal}, P., {Hirashita}, H., \& {Ferrara}, A. 2010, \mnras, 403, 620

\bibitem[{{Draine}(2009)}]{Draine09}
{Draine}, B.~T. 2009, in Astronomical Society of the Pacific Conference Series,
  Vol. 414, Cosmic Dust - Near and Far, ed. T.~{Henning}, E.~{Gr{\"u}n}, \&
  J.~{Steinacker}, 453

\bibitem[{{Dulieu} {et~al.}(2010){Dulieu}, {Amiaud}, {Congiu}, {Fillion},
  {Matar}, {Momeni}, {Pirronello}, \& {Lemaire}}]{Dulieu10}
{Dulieu}, F., {Amiaud}, L., {Congiu}, E., {Fillion}, J.-H., {Matar}, E.,
  {Momeni}, A., {Pirronello}, V., \& {Lemaire}, J.~L. 2010, \aap, 512, A30

\bibitem[{{Fayolle} {et~al.}(2013){Fayolle}, {Bertin}, {Romanzin}, {Poderoso},
  {Philippe}, {Michaut}, {Jeseck}, {Linnartz}, {{\"O}berg}, \&
  {Fillion}}]{Fayolle13}
{Fayolle}, E.~C., {et~al.} 2013, \aap, 556, A122

\bibitem[{{Field} {et~al.}(1969){Field}, {Goldsmith}, \& {Habing}}]{Field69}
{Field}, G.~B., {Goldsmith}, D.~W., \& {Habing}, H.~J. 1969, \apjl, 155, L149

\bibitem[{{Fraser} {et~al.}(2001){Fraser}, {Collings}, {McCoustra}, \&
  {Williams}}]{Fraser01}
{Fraser}, H.~J., {Collings}, M.~P., {McCoustra}, M.~R.~S., \& {Williams}, D.~A.
  2001, \mnras, 327, 1165

\bibitem[{{Gall} {et~al.}(2011){Gall}, {Hjorth}, \& {Andersen}}]{Gall11}
{Gall}, C., {Hjorth}, J., \& {Andersen}, A.~C. 2011, \aapr, 19, 43

\bibitem[{{Gall} {et~al.}(2014){Gall}, {Hjorth}, {Watson}, {Dwek}, {Maund},
  {Fox}, {Leloudas}, {Malesani}, \& {Day-Jones}}]{Gall14}
{Gall}, C., {et~al.} 2014, \nat, 511, 326

\bibitem[{{Hama} {et~al.}(2012){Hama}, {Kuwahata}, {Watanabe}, {Kouchi},
  {Kimura}, {Chigai}, \& {Pirronello}}]{Hama12}
{Hama}, T., {Kuwahata}, K., {Watanabe}, N., {Kouchi}, A., {Kimura}, Y.,
  {Chigai}, T., \& {Pirronello}, V. 2012, \apj, 757, 185

\bibitem[{{Hasegawa} \& {Herbst}(1993)}]{Hasegawa93}
{Hasegawa}, T.~I., \& {Herbst}, E. 1993, \mnras, 261, 83

\bibitem[{{Hasegawa} {et~al.}(1992){Hasegawa}, {Herbst}, \&
  {Leung}}]{Hasegawa92}
{Hasegawa}, T.~I., {Herbst}, E., \& {Leung}, C.~M. 1992, \apjs, 82, 167

\bibitem[{{He} {et~al.}(2016){He}, {Acharyya}, \& {Vidali}}]{He16}
{He}, J., {Acharyya}, K., \& {Vidali}, G. 2016, \apj, 823, 56

\bibitem[{{Hirashita} \& {Ferrara}(2002)}]{Hirashita02}
{Hirashita}, H., \& {Ferrara}, A. 2002, \mnras, 337, 921

\bibitem[{{Hirashita} \& {Voshchinnikov}(2014)}]{Hirashita14a}
{Hirashita}, H., \& {Voshchinnikov}, N.~V. 2014, \mnras, 437, 1636

\bibitem[{{Hornek{\ae}r} {et~al.}(2005){Hornek{\ae}r}, {Baurichter},
  {Petrunin}, {Luntz}, {Kay}, \& {Al-Halabi}}]{Hornekaer05}
{Hornek{\ae}r}, L., {Baurichter}, A., {Petrunin}, V.~V., {Luntz}, A.~C., {Kay},
  B.~D., \& {Al-Halabi}, A. 2005, \jcp, 122, 124701

\bibitem[{{Indebetouw} {et~al.}(2014){Indebetouw}, {Matsuura}, {Dwek},
  {Zanardo}, {Barlow}, {Baes}, {Bouchet}, {Burrows}, {Chevalier}, {Clayton},
  {Fransson}, {Gaensler}, {Kirshner}, {Laki{\'c}evi{\'c}}, {Long}, {Lundqvist},
  {Mart{\'{\i}}-Vidal}, {Marcaide}, {McCray}, {Meixner}, {Ng}, {Park},
  {Sonneborn}, {Staveley-Smith}, {Vlahakis}, \& {van Loon}}]{Indebetouw14}
{Indebetouw}, R., {et~al.} 2014, \apjl, 782, L2

\bibitem[{{Jones} \& {Nuth}(2011)}]{Jones11}
{Jones}, A.~P., \& {Nuth}, J.~A. 2011, \aap, 530, A44

\bibitem[{{Krasnokutski} {et~al.}(2014){Krasnokutski}, {Rouille}, {Jager},
  {Huisken}, {Zhukovska}, \& {Henning}}]{Krasnokutski14}
{Krasnokutski}, S.~A., {Rouille}, G., {Jager}, C., {Huisken}, F., {Zhukovska},
  S., \& {Henning}, T. 2014, \apj, 782, 15

\bibitem[{{Leger} {et~al.}(1985){Leger}, {Jura}, \& {Omont}}]{Leger85}
{Leger}, A., {Jura}, M., \& {Omont}, A. 1985, \aa, 144, 147

\bibitem[{{Mancini} {et~al.}(2015){Mancini}, {Schneider}, {Graziani},
  {Valiante}, {Dayal}, {Maio}, {Ciardi}, \& {Hunt}}]{Mancini15}
{Mancini}, M., {Schneider}, R., {Graziani}, L., {Valiante}, R., {Dayal}, P.,
  {Maio}, U., {Ciardi}, B., \& {Hunt}, L.~K. 2015, \mnras, 451, L70

\bibitem[{{Matar} {et~al.}(2008){Matar}, {Congiu}, {Dulieu}, {Momeni}, \&
  {Lemaire}}]{Matar08}
{Matar}, E., {Congiu}, E., {Dulieu}, F., {Momeni}, A., \& {Lemaire}, J.~L.
  2008, \aap, 492, L17

\bibitem[{{Matsuura} {et~al.}(2011){Matsuura}, {Dwek}, {Meixner}, {Otsuka},
  {Babler}, {Barlow}, {Roman-Duval}, {Engelbracht}, {Sandstrom},
  {Laki{\'c}evi{\'c}}, {van Loon}, {Sonneborn}, {Clayton}, {Long}, {Lundqvist},
  {Nozawa}, {Gordon}, {Hony}, {Panuzzo}, {Okumura}, {Misselt}, {Montiel}, \&
  {Sauvage}}]{Matsuura11}
{Matsuura}, M., {et~al.} 2011, Science, 333, 1258

\bibitem[{{Micha{\l}owski} {et~al.}(2010){Micha{\l}owski}, {Murphy}, {Hjorth},
  {Watson}, {Gall}, \& {Dunlop}}]{Michalowski10}
{Micha{\l}owski}, M.~J., {Murphy}, E.~J., {Hjorth}, J., {Watson}, D., {Gall},
  C., \& {Dunlop}, J.~S. 2010, \aap, 522, A15

\bibitem[{{Minissale} {et~al.}(2016){Minissale}, {Congiu}, \&
  {Dulieu}}]{Minissale16}
{Minissale}, M., {Congiu}, E., \& {Dulieu}, F. 2016, \aap, 585, A146

\bibitem[{{Nozawa} {et~al.}(2007){Nozawa}, {Kozasa}, {Habe}, {Dwek}, {Umeda},
  {Tominaga}, {Maeda}, \& {Nomoto}}]{Nozawa07}
{Nozawa}, T., {Kozasa}, T., {Habe}, A., {Dwek}, E., {Umeda}, H., {Tominaga},
  N., {Maeda}, K., \& {Nomoto}, K. 2007, \apj, 666, 955

\bibitem[{{{\"O}berg} {et~al.}(2009){{\"O}berg}, {Linnartz}, {Visser}, \& {van
  Dishoeck}}]{Oberg09}
{{\"O}berg}, K.~I., {Linnartz}, H., {Visser}, R., \& {van Dishoeck}, E.~F.
  2009, \apj, 693, 1209

\bibitem[{{Planck Collaboration} {et~al.}(2015){Planck Collaboration}, {Ade},
  {Aghanim}, {Arnaud}, {Ashdown}, {Aumont}, {Baccigalupi}, {Banday},
  {Barreiro}, {Bartlett}, \& et~al.}]{Ade15}
{Planck Collaboration} {et~al.} 2015, ArXiv e-prints

\bibitem[{{Shannon} {et~al.}(2013){Shannon}, {Blitz}, {Goddard}, \&
  {Heard}}]{Shannon13}
{Shannon}, R.~J., {Blitz}, M.~A., {Goddard}, A., \& {Heard}, D.~E. 2013, Nature
  Chemistry, 5, 745

\bibitem[{{Shen} {et~al.}(2004){Shen}, {Greenberg}, {Schutte}, \& {van
  Dishoeck}}]{Shen04}
{Shen}, C.~J., {Greenberg}, J.~M., {Schutte}, W.~A., \& {van Dishoeck}, E.~F.
  2004, \aa, 415, 203

\bibitem[{{Spitzer}(1978)}]{Spitzer78}
{Spitzer}, L. 1978, {Physical processes in the interstellar medium}

\bibitem[{{Stutz} {et~al.}(2010){Stutz}, {Launhardt}, {Linz}, {Krause},
  {Henning}, {Kainulainen}, {Nielbock}, {Steinacker}, \& {Andr{\'e}}}]{Stutz10}
{Stutz}, A., {et~al.} 2010, \aap, 518, L87

\bibitem[{{Taquet} {et~al.}(2012){Taquet}, {Ceccarelli}, \&
  {Kahane}}]{Taquet12}
{Taquet}, V., {Ceccarelli}, C., \& {Kahane}, C. 2012, \aa, 538, 42

\bibitem[{{Todini} \& {Ferrara}(2001)}]{Todini01}
{Todini}, P., \& {Ferrara}, A. 2001, \mnras, 325, 726

\bibitem[{{Umebayashi} \& {Nakano}(1980)}]{Umebayashi80}
{Umebayashi}, T., \& {Nakano}, T. 1980, \pasj, 32, 405

\bibitem[{{Vallini} {et~al.}(2013){Vallini}, {Gallerani}, {Ferrara}, \&
  {Baek}}]{Vallini13}
{Vallini}, L., {Gallerani}, S., {Ferrara}, A., \& {Baek}, S. 2013, \mnras, 433,
  1567

\bibitem[{{Weingartner} \& {Draine}(2001)}]{Weingartner01}
{Weingartner}, J.~C., \& {Draine}, B.~T. 2001, \apj, 548, 296

\bibitem[{{Wolfire} {et~al.}(2003){Wolfire}, {McKee}, {Hollenbach}, \&
  {Tielens}}]{Wolfire03}
{Wolfire}, M.~G., {McKee}, C.~F., {Hollenbach}, D., \& {Tielens}, A.~G.~G.~M.
  2003, \apj, 587, 278

\end{thebibliography}

\newpage 
\label{lastpage} 
\end{document}